**Conversion of Heat into Charge Current by the Spin Wave Anomalous Nernst Effect**

M. Mizuguchi[1,2,3], K. Hasegawa[1], J. Ohe[2,3], M. Tsujikawa[3,4], M. Shirai[3,4] and K. Takanashi[1,3]

[1]*Institute for Materials Research, Tohoku University, Sendai 980-8577, Japan*

[2]*CREST, Japan Science and Technology Agency, Sanbancho, Tokyo 102-0075, Japan*

[3]*Center for Spintronics Research Network, Tohoku University, Sendai 980-8577, Japan*

[4]*Department of Physics, Toho University, 2-2-1 Miyama, Funabashi, Japan*

[5]*Research Institute of Electrical Communication, Tohoku University, Sendai 980-8577, Japan*

Study of the various aspects of spin conversion is an intriguing research field since an angular momentum flows of electrons, spin waves, circularly polarized light and mechanical motion of crystals have been recognized as carriers of spin current. In addition, phenomena relating to spin conversion between different physical flows are anticipated to cultivate renewed energy conversion schemes that could serve as a useful energy sources for a next society embracing the Internet of Things. A novel process of spin conversion from a temperature gradient to a transverse voltage is addressed in this paper, viz. the anomalous Nernst effect (ANE) in a ferromagnetic metal. We report that an additional voltage is superposed on the conventional anomalous Nernst voltage in FePt crystalline thin films. The dynamics of the local magnetization is modulated by the heat current and excites spin waves. These generate a conduction electron spin current via s-d coupling, which flows along the temperature gradient, and the spin current is converted to a Nernst voltage by the inverse spin Hall effect. This spin conversion would be observable at any position of a single ferromagnet without any detection materials but this requires a ferromagnetic metal with strong spin-orbit coupling such as FePt alloy. To substantiate our ideas, ANE and anomalous Hall effect (AHE) in $L1_0$-ordered FePt thin films with different uniaxial magnetic anisotropies ($K_u$) were investigated. The ANE decreased with $K_u$, whereas the AHE increased with increasing $K_u$. The measurement temperature dependence of both effects clarifies that the Nernst conductivity ($\alpha_{xy}$) is well fitted by the Mott equation below 100 K, whereas $\alpha_{xy}$ is relatively enhanced above 100 K. This indicates that a certain number of magnons is required such that spin wave excitation can be converted to a pure conduction electron spin current by the temperature gradient to enhance the ANE in FePt.

Spin-caloritronics deals with the coupled heat, spin current and currents and is currently studied energetically.[1-7] Generation of a spin current by a temperature gradient is widely studied in the spin Seebeck effect, which can be exploited as a thermo-magnetic conversion system.[8-10] However, the spin Seebeck effect needs a spin detection material with strong spin-orbit coupling such as Pt, in which the thermally generated spin current can be converted into a voltage only within a very short spin diffusion length. A spin conversion process which occurs in the bulk of a magnet is strongly desired for an efficient thermomagnetic conversion device. For that reason, we have been investigating the anomalous Nernst effect (ANE), which has been known as a common thermomagnetic effect for a long time.[11] In a previous paper,[12] we reported ANE measurements of an $L1_0$-ordered epitaxial FePt thin film for studying thermomagnetic effects in ordered alloys, and we evaluated the anomalous Nernst coefficient and the anomalous Nernst angle of an FePt thin film. We also reported a material dependence of the ANE in various ordered alloys and clarified that the ANE has a tendency to increase monotonously with the uniaxial magnetic anisotropy ($K_u$) at room temperature.[13]

On the other hand, the anomalous Hall effect (AHE) in ferromagnetic metals, is also well known. It has been reported that the ANE and AHE are related by the phenomenological Mott relation;

$$\alpha_{xy} = T\,\pi^2 k_B^2\,/3e \times (\partial\rho_{yx}/\partial E)_{EF}, \qquad\qquad (1)$$

where $E$, $\alpha_{xy}$, $\rho_{yx}$, and $(\partial\rho_{yx}/\partial E)_{EF}$ express the energy, the Nernst conductivity, the Hall conductivity, the Hall resistivity, and the partial differential of the Hall resistivity at the Fermi energy, respectively. Mott relations have been studied in transition metal oxides and ferromagnetic semiconductors.[16] However, the comparisons were limited only to a very small temperature range and there have been few reports on ferromagnetic metals and alloys.[17] It is essential to investigate the relation between ANE, AHE and magnetic anisotropy to unravel the mechanism of the ANE because the spin-orbit coupling plays a key role in all ANE, AHE and magnetic anisotropy.

The $K_u$ values of FePt thin films depend on the $L1_0$-order parameter that can be controlled by the substrate temperature during sputtering.[18] From this point of view, this Letter describes our investigation of the ANE and AHE in $L1_0$-ordered FePt thin films with different $K_u$. Measurement temperature dependences of the two effects were also measured and experiments were compared to the Mott relation. It has been found that spin wave excitation by the temperature gradient is converted to pure conduction electron spin current and eventually enhances the ANE in FePt. This conversion of spin current occurs anywhere in a ferromagnet and needs any kind of the spin detection pads as in the spin Seebeck effect.[3]

Figure 1 (a) shows the X-ray diffraction (XRD) patterns of FePt thin films fabricated at various substrate temperatures. Superlattice peaks of FePt 001 and 003 were observed for all the samples in addition to fundamental peaks of FePt 002 and 004. This confirms that all the FePt thin films form epitaxial $L1_0$-ordered structure on MgO substrates, and the order parameter was varied by changing the substrate temperature. Figure 1 (b) shows in-plane and out-of-plane magnetization curves of the FePt thin films measured at RT. It was revealed that magnetic anisotropy increased with increasing the substrate temperature. Uniaxial magnetic anisotropy energy ($K_u$) was estimated from the enclosed area between in-plane and out-of-plane magnetization curves. $K_u$ simply changed depending on the order parameter of $L1_0$-FePt.

Distinct electromotive forces were observed in ANE and AHE measurements at RT for all FePt films with various $K_u$. The voltage of ANE and AHE changed with a hysteresis depending on the magnetic field, and the magnetic fields where the voltage crossed 0 V and saturated almost coincided with the coercivity and the saturation field, respectively, in the out-of-plane magnetization curves. Figure 2 summarizes the $K_u$ dependence of ANE and AHE of FePt thin films. $\alpha_{xy}$ and Hall conductivity ($\sigma_{xy}$) are defined as following equations, respectively:

$$\alpha_{xy} = \sigma_{xx} S_{xy} + \sigma_{xy} S_{xx}, \qquad (2)$$

$$\sigma_{xy} = \rho_{Hall} / \rho_{xx}^2, \tag{3}$$

where $S_{xx}$, $S_{xy}$, and $\rho_{xx}$ express the Seebeck coefficient, the transverse Seebeck coefficient, and the electrical resistivity, respectively.

$S_{xy}$ is derived from the following equation.

$$S_{xy} = E_y / \nabla T_x, \tag{4}$$

where $E_y$ and $\nabla T_x$ express the Nernst electric field and the temperature gradient, respectively. $\sigma_{xy}$ increases with increasing $K_u$. This tendency is easy to understand because the strength of the spin-orbit coupling inside FePt crystals increases with the magnetic anisotropy. On the other hand, $\alpha_{xy}$ monotonously decreases with $K_u$. This tendency implies that additional effects contribute to the ANE besides the spin-orbit coupling.

To clarify the origin of the additional contribution, the temperature dependence of $S_{xy}$ was investigated. Figure 3 (a) shows the $S_{xy}$ in $L1_0$-FePt fabricated at the substrate temperature of 500˚C ($K_u = 2.6 \times 10^7$ erg/cc) measured at several temperatures. It is revealed that the hysteresis loop of $S_{xy}$ measured at 300 K shrinks gradually and vanishes eventually at 10 K. This means that the ANE increases with the temperature, and this tendency is common to other FePt films fabricated at different substrate temperatures. The temperature dependence of AHE was also measured in FePt thin films, and $\rho_{Hall}$ increases with temperature for all the FePt films. The Seebeck coefficient ($S_{xx}$) and the transverse Seebeck coefficient ($S_{xy}$) in $L1_0$-FePt thin films with different $K_u$ as a function of temperature are plotted in Fig. 3(b) and (c), respectively. No clear difference in $S_{xx}$ was seen for three films whereas they revealed clear differences in $S_{xy}$. As seen in this figure, $S_{xy}$ increases with decreasing $K_u$, particularly, over 100 K. The temperature dependence of $\alpha_{xy}$ for FePt thin films fabricated at the substrate temperature of 500˚C is plotted in Fig. 3 (d) by using equation (2). We observe a remarkable increase of $\alpha_{xy}$ with temperature starting at ca. 100 K. The low temperature experimental values were fitted by the Mott equation involving the prefactor as a parameter, which is derived from the equation (1) as follows:

$$\alpha_{xy} = \rho_{xy} / \rho^2_{xx} \cdot \{T \pi^2 k^2_B \, d \, \lambda /(3e \, \lambda \, d \, \varepsilon )\text{-} (n\text{-}2) \, S_{xx}\}, \qquad (5)$$

where $\lambda$ and $n$ are the fitting parameters depending on the Fermi energy of conduction electrons and the power factor relations between $\rho_{xx}$ and $\rho_{xy}$, respectively.[16] From the experiment of AHE, we found the relation of $\rho_{xy} \propto \rho^2_{xx}$, which implies that the main mechanism of the AHE is intrinsic, thus $n = 2$ was used in the fitting. From a fitting of ANE using the equation (5) with free parameters of $\lambda$ and $n$, $n$ was also determined to be 2. The solid line in Fig. 3 (d) shows the fitted line for $\alpha_{xy}$. It was found that $\alpha_{xy}$ was well fitted by the Mott equation below 100 K, whereas $\alpha_{xy}$ apparently enhanced more than fitted line over 100 K. The enhancement of the ANE can be explained by the contribution from the conduction electron spin current induced by the magnetization dynamics in the presence of the temperature gradient. We consider the s-d coupling between the conduction electrons and the local magnetization. The Hamiltonian of the conduction electron is expressed as follows:

$$H = \frac{p^2}{2m} - J_{sd}\boldsymbol{S} \cdot \boldsymbol{M} \qquad (6),$$

where $J_{sd}$ is the s-d coupling energy, $S$ is the spin matrix, and $M$ is the local magnetic moment. Not only the mass flow of electrons but also collective excitation of magnetic moments, i.e. spin waves, carry a heat current.

Figure 4(a) depicts a schematic image of the generation of conduction electron spin currents by the temperature gradient. The magnetization dynamics obeys the Landau-Lifshitz-Gilbert equation

$$\frac{\partial \mathbf{M}}{\partial t} = -\gamma(\mathbf{M} \times \mathbf{H}_{eff}) + \alpha\left(\mathbf{M} \times \frac{\partial \mathbf{M}}{\partial t}\right) \qquad (7)$$

where $\gamma$ is the gyromagnetic ratio and $\alpha$ is the Gilbert damping coefficient. The effective magnetic field ($\mathbf{H}_{eff}$) includes the effect of the exchange coupling, anisotropic energy, dipole-dipole interaction, and random magnetic field due to the finite temperature. In the presence of temperature gradient, the stochastic magnetic field generates a gradient in the magnetization. This magnetization dynamics is a source of conduction electron spins through the s-d coupling. The source of the spin is expressed as:

$$d\mathbf{S}/dt \propto \mathbf{M} \times \frac{\partial \mathbf{M}}{\partial t} \qquad (8)$$

because a conduction electron spin accumulation is generated proportional to the magnetization damping. The spatial gradient of the conduction electron spin accumulation generates the conduction electron spin current along the direction of the temperature gradient, and is converted to a transverse charge current by the inverse spin Hall effect.[19] Because the pure spin current does not contribute to the charge transport in the direction of the temperature gradient, the experimental results does not show any enhancement of the longitudinal Seebeck coefficient as shown in Fig. 3(b). In order to confirm this mechanism, we carried out micromagnetic simulations of the magnetization dynamics. Literature values were employed for $J_{sd}$ and $\alpha$. Figure 4 (b) shows the spatial distribution of the conduction electron spin source. When the temperature gradient is applied, the net spin source of the conduction electron is calculated by the difference between the spin source of the uniform temperature and that of the temperature gradient. The conduction electron spin source becomes larger in the high temperature region. We calculate the ANE signal from the pure spin current in FePt as shown in Fig. 4 (c) and 4 (d). It is assumed that the ANE signal is proportional to the total conduction electron spin source and they are integrated over the sample. ANE signal of FePt as a function of $K_u$ was calculated with the temperature gradient of 10 K/6 μm as shown in Fig. 4(c). ANE signal monotonously decreased with $K_u$. This tendency agrees well to the experiment which is also shown in the figure. The spin source expressed in Eq.(8) becomes larger when the fluctuation of the magnetization is large. From the view of the magnetization dynamics, the spin source is expressed as the divergence of the spin-wave spin current. With increasing $K_u$, the generation of the spin-wave tends to decrease, thus it is considered that the ANE signal from the pure spin current decreases.

Temperature dependence of the ANE signal was also calculated by the same method as Fig. 4(c) as shown in Figure 4 (d). The ANE signal increases with temperature and the contribution of the conduction electron spin current saturates over 100 K. From this simulation, we conclude that the magnetic order is excited significantly only over 100 K for this FePt which agrees with the $\alpha_{xy}$ enhancement beyond the Mott relation. It can be considered that this spin wave excitation by the temperature gradient converted to the pure conduction electron spin current and eventually enhanced the ANE in FePt.

This novel process is expected to pave a new way for renewed energy conversion systems which will serve a useful energy sources in our daily life. The expectations of realization of ANE-based thermoelectric devices strongly rise by employing this novel spin-conversion process.

The authors are grateful to Y. Sakuraba of NIMS for their help for low-temperature measurements. The authors are grateful to K. Uchida, E. Saitoh, J. Barker, K. Nomura, and G.E.W. Bauer of Tohoku University for fruitful discussions. This research was partly supported by a Grant-in-Aid for Scientific Research (S) (25220910) from Japan Society for the Promotion of Science and CREST (Grant No. JPMJCR1524) from Japan Science and Technology Agency.

**Figure 1.** (a) The XRD patterns of FePt thin films fabricated at various substrate temperatures taken at RT. The order parameter of each film is indicated. (b) In-plane and out-of-plane magnetization curves of the same FePt thin films as in (a) measured at RT.

**Figure 2.** The $K_u$ dependence of Nernst conductivity ($\alpha_{xy}$) and Hall conductivity ($\sigma_{xy}$) of FePt thin films.

**Figure 3.** (a) The transverse Seebeck coefficient ($S_{xy}$) in $L1_0$-FePt fabricated at the substrate temperature of 500˚C ($K_u = 2.6 \times 10^7$ erg/cc) measured at several temperatures. (b) The Seebeck coefficient ($S_{xx}$) in $L1_0$-FePt as a function of temperature. (c) The transverse Seebeck coefficient ($S_{xy}$) in $L1_0$-FePt as a function of temperature. (d) The temperature dependence of $\alpha_{xy}$ for FePt thin films fabricated at the substrate temperature of 500˚C (closed dots) and the fitted line for $\alpha_{xy}$ by the Mott equation (solid line).

**Figure 4.** (a) Schematic image of generation of spin current by temperature gradient in FePt. (b) Calculated spin accumulation generated by a temperature difference of 10 K (in arbitrary units). (c) Calculated ANE signal of FePt as a function of $K_u$ with the temperature difference of 10 K. Experimental values are also shown. (d) Calculated ANE signal of FePt as a function of temperature with the temperature difference of 10 K.

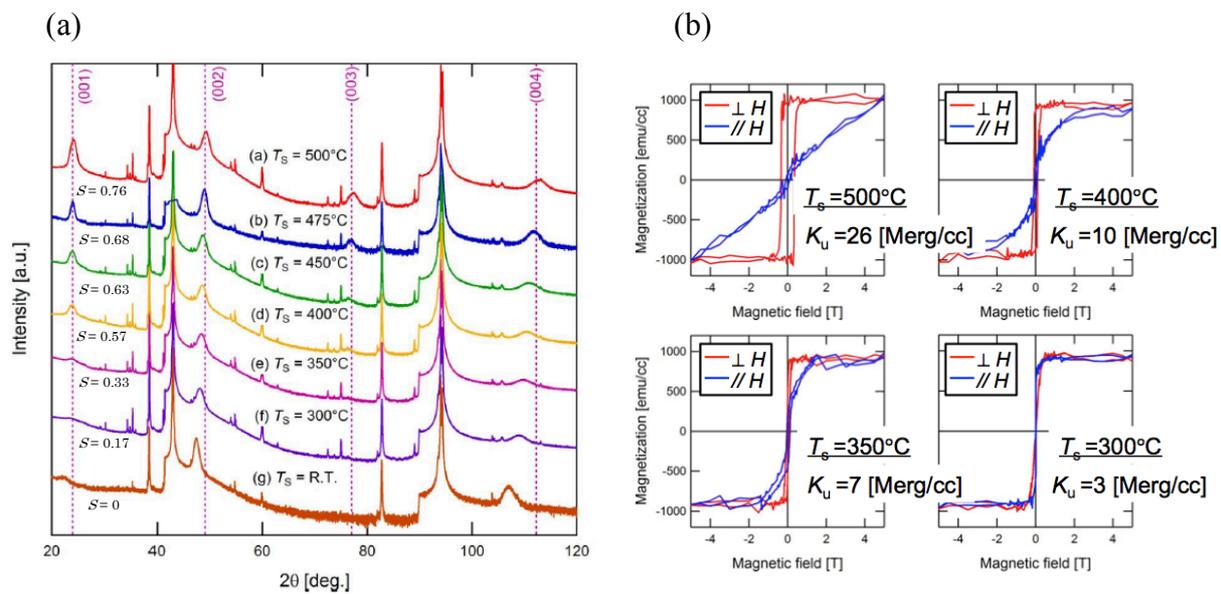

Figure 1

M. Mizuguchi *et al*.

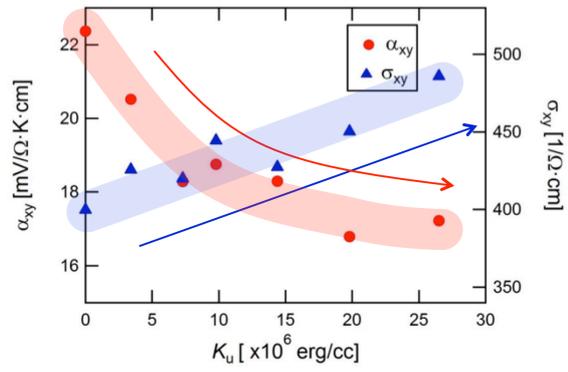

Figure 2

M. Mizuguchi *et al*.

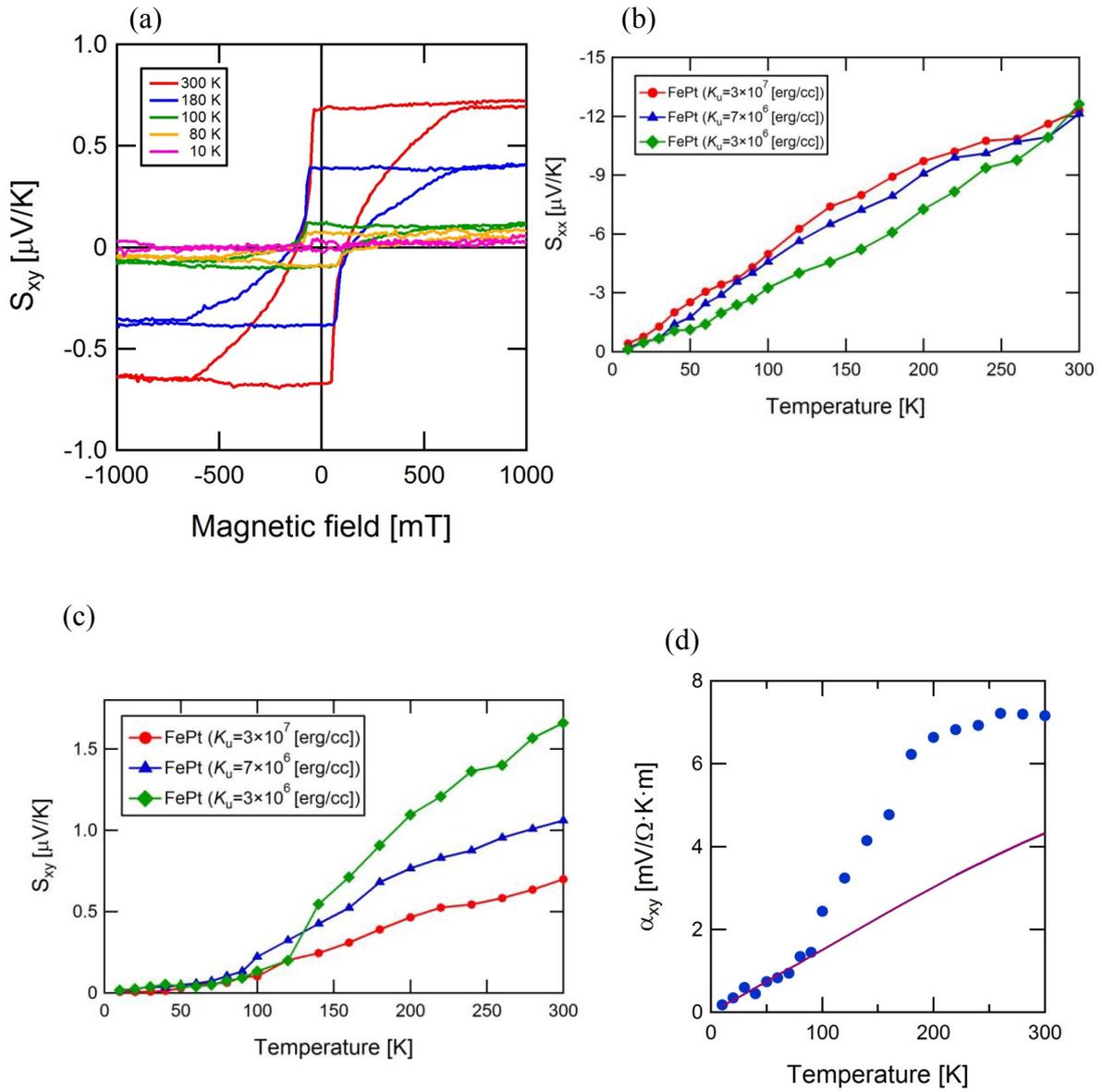

Figure 3

M. Mizuguchi *et al*.

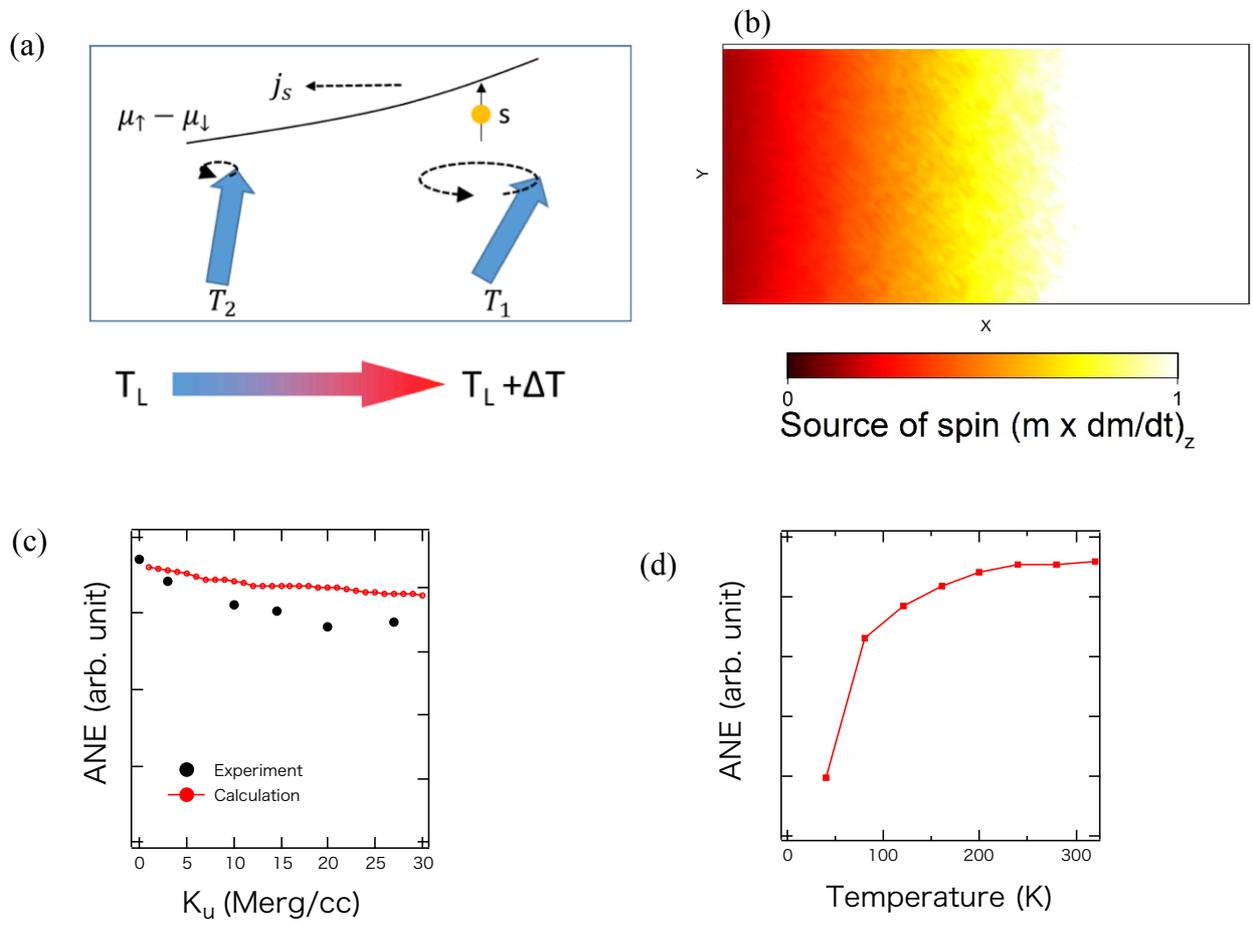

(a)

$j_s$

$\mu_\uparrow - \mu_\downarrow$

s

$T_2$

$T_1$

$T_L$ → $T_L + \Delta T$

(b)

Y

X

Source of spin (m × dm/dt)$_z$

0      1

(c)

ANE (arb. unit)

$K_u$ (Merg/cc)

0   5   10   15   20   25   30

● Experiment
● Calculation

(d)

ANE (arb. unit)

Temperature (K)

0   50   100   150   200   250   300

Figure 4

M. Mizuguchi *et al*.